\documentclass[twocolumn,groupedaddress,aps,pra,groupedaddress,amsmath,amssymb,showpacs]{revtex4-1}
\usepackage{graphicx,amsmath,amssymb,epsfig,color}
\usepackage{dcolumn}
\usepackage{bm}
\usepackage[colorlinks,	linkcolor=red,	anchorcolor=blue,citecolor=blue,bookmarksnumbered]{hyperref}

\begin{document}
\preprint{}

\title{Stable single light bullets and vortices and their active control in cold Rydberg gases}
\author{Zhengyang Bai$^{1 }$, Weibin Li$^{2,4}$, and Guoxiang Huang$^{1,3}$\footnote{gxhuang@phy.ecnu.edu.cn} }
\affiliation{
             $^1$State Key Laboratory of Precision Spectroscopy,
                 East China Normal University, Shanghai 200062, China\\
             $^2$School of Physics and Astronomy, University of Nottingham, Nottingham, NG7 2RD, UK\\
             $^3$NYU-ECNU Joint Institute of Physics at NYU-Shanghai, Shanghai 200062, China\\
             $^4$Centre for the Mathematics and Theoretical Physics of Quantum Non-equilibrium Systems, University of Nottingham, Nottingham, NG7 2RD, UK\\
             }
\date{\today}

\begin{abstract}

Realizing single light bullets and vortices that are stable in high dimensions is a long-standing goal in the study of nonlinear optical physics. On the other hand, the storage and retrieval of such stable high dimensional optical pulses may offer a variety of applications. Here we present a scheme to generate such optical pulses in a cold Rydberg atomic gas. By virtue of electromagnetically induced transparency, strong, long-range atom-atom interaction in Rydberg states is mapped to light fields, resulting in a giant, fast-responding nonlocal Kerr nonlinearity and the formation of light bullets and vortices carrying orbital angular momenta,
which have extremely low generation power, very slow propagation velocity, and can stably propagate, with the stability provided by the combination of local and the nonlocal Kerr nonlinearities. We demonstrate that the light bullets and vortices obtained can be stored and retrieved in the system with high efficiency and fidelity. Our study provides a new route for manipulating high-dimensional nonlinear optical processes via the controlled optical nonlinearities in cold Rydberg gases.


\end{abstract}

\maketitle

\section{INTRODUCTION}{\label{Sec1}}

High-dimensional spatiotemporal optical solitons, alias light bullets (LBs)~\cite{Silberberg}, are solitary nonlinear waves localized in $n$-spatial dimensions and one time axis [($m$+1)D; $m=1,2,3$]. In recent years, the study of LBs has attracted intensive theoretical and experimental interests~\cite{Kivshar} because of their rich nonlinear physics and technological applications~\cite{Malomed,Mihalache,Belic,Miha,Mateo,Karta}. Experimental signatures of LBs in different types of optical media, such as (2+1)D LBs in LiIO$_3$ crystals~\cite{Liu} and  quasi-(3+1)D LBs in waveguide arrays~\cite{Minardi,Eile}, have been reported. Recently, trains of (3+1)D dark solitons have been observed in a photo-refractive material by Lahav {\it et al}.~\cite{Lah}. These LBs, however, generally suffer severe instability, which typically can propagate to only a few diffraction lengths. To levitate the rapid distortion of LBs, short pulses (in the order of femtosecond) with high laser powers were commonly used experimentally. Despite fast experimental progresses, it remains a challenge to create stable single LBs, as they can only be realized by carefully balancing their dispersion, diffraction and optical nonlinearities.

Theoretical efforts have attempted to study the formation of LBs with different mechanisms. A commonly used approach for creating stable (3+1)D LBs is to exploit local and nonlocal optical nonlinearities with quite different response times. These nonlinearities have been examined, e.g., in nematic liquid crystals~\cite{Bur,Pec1,Pec2} and lead glass~\cite{Gur}, in which the nonlocal nonlinearity (resulted from the reorientation~\cite{Bur,Pec1,Pec2} or thermal motion~\cite{Gur} of molecules) has a very long response time (typically in the order of second or even longer), while the local nonlinearity (resulted from the electronic Kerr effect) has a very short response time (in the order of femtosecond). Due to the mismatch of the response-time scales and the fast propagation velocity ($\sim c$) of optical pulses, stable LBs can only be realized in the form of pulse-train solitons (not single LBs)~\cite{Lah}. A different approach is based on electromagnetically induced transparency (EIT) in atomic gases~\cite{FIM}, in which optical absorption can be largely suppressed due to quantum interference. Together with Kerr nonlinearities~\cite{Hau, Pri} induced by resonant laser fields, it has been shown that stable (1+1)D temporal~\cite{Wu,huang,Chen} and spatial~\cite{Hong,Michinel} optical solitons can form. Though promising, the local nature of the Kerr nonlinearity doesn't support stable LBs, which suffer unavoidable catastrophic collapse during propagation.

On the other hand, recent theoretical~\cite{Friedler}  and experimental~\cite{Moha} studies revealed that strong and long-range optical nonlinearities can be built with Rydberg atomic gases~\cite{REITreview,Fir,Mur}, which are in electronically high-lying states with large principal quantum number $n$~\cite{Gallagher}. Large dipole moments in Rydberg states render a strong, long-range Rydberg-Rydberg interaction between remote atoms. Rydberg interactions find applications in quantum information, precision measurement, quantum simulation  based on Rydberg atoms~\cite{Saffman,Low,REITreview,Fir,Mur}. Importantly, the Rydberg-Rydberg interaction can be mapped to a nonlocal optical nonlinearity through EIT, which is strong even at the single photon level~\cite{REITreview,Fir,Mur}. Long lifetimes (in the order of tens of microsecond) guarantee that the induced photon-photon interaction is largely coherent during light propagation~\cite{Gor0,BingHe,LiuYang}. This provides a new platform to study quantum nonlinear optics~\cite{Fir} and develop new photon devices, such as single-photon switches and transistors~\cite{Baur,Gorn1,Gorn2}, quantum memories and phase gates~\cite{Maxwell2013,Distante,LiL,Busche,Tia2,Liang}.

In this work, we present a scheme for the generation and storage of stable (3+1)D LBs in highly tunable cold gases of Rydberg atoms. A key element is the co-existence of giant nonlocal and local optical nonlinearities. The former features a fast (sub-microsecond)  response~\cite{Zhang}, which is complemented by the latter~\cite{thomas,Gullans}, whose response is relative slower (in the order of microsecond). In conjunction with  tunable dispersion and diffraction, this allows us to precisely control dynamics of LBs. To go beyond the commonly used mean-field theory~\cite{Sche,Sev,Stan,Bai}, we derive a nonlinear (3+1)D light propagation equation taking into account of one-body and two-body correlations.
We show that the {\it synergetic} nonlocal and local optical nonlinearities in the system permit us to obtain stable single (3+1)D LBs as well as LBs that carry definite orbital angular momentum [i.e. light vortices (LVs)\,]~\cite{Andrews,Veissier,Nicolas,Parigi,Oliveira,ding}. It is revealed that the stability of the single LBs and LVs is achieved via a {\it two-step self-trapping mechanism}:  First, the transverse self-trapping of an input high-dimensional laser pulse is quickly reached through the balance between the diffraction and the fast nonlocal nonlinearity; then the laser pulse is further self-trapped in the longitudinal (propagation) direction through the interplay between the dispersion and the slow local nonlinearity. As a result, stable bright (3+1)D LBs can be generated with ultraslow propagation velocity, extremely low light power, and narrow bandwidth. More importantly, based on the active character of the system, we can manipulate the LBs and LVs by tuning system parameters, and can even store and retrieve LBs and LVs in the Rydberg medium with high efficiency and fidelity.

The paper is arranged as follows. In Sec.~\ref{sec2}, we describe the physical model of the Rydberg-EIT system. In Sec.~\ref{sec3}, we derive the (3+1)D nonlocal nonlinear envelope equation based on the multiple-scale  method.
In Sec.~\ref{sec4}, we discuss the physical property of the nonlocal and local Kerr nonlinearities and present LB and LV solutions in various regimes. Based on numerical calculations,  we illustrate in Sec.~\ref{sec5} that LBs can be used as optical memories such that they can be dynamically stored and read out in the Rydberg medium. Finally, in the last section (Sec.~\ref{sec6}) we summarize the main results obtained in this work.

\section{Model}\label{sec2}

\begin{figure}
\centering
\includegraphics[width=0.48\textwidth]{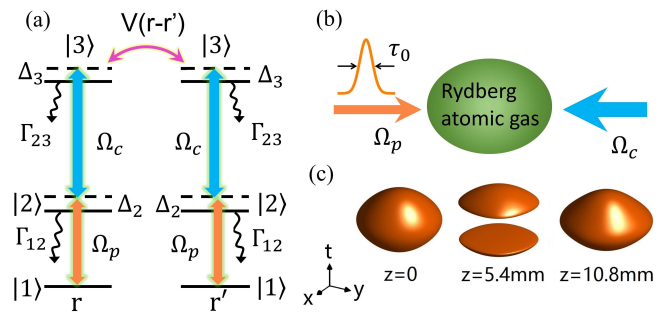}\\
\caption{\footnotesize(Color online) (a)~EIT level scheme, where the ground state $|1\rangle$, intermediate state $|2\rangle$, and Rydberg state $|3\rangle$ are respectively driven by a pulsed probe field (with pulse duration $\tau_0$) and a strong control field. State $|2\rangle$ has a large spontaneous decay rate $\Gamma_{12}\sim$ MHz. The weak decay $\Gamma_{23}\sim$ kHz from $|3\rangle$ to $|2\rangle$ is also taken into account. The van der Waals interaction $\hbar V(\bf{r}-\bf{r}')$  between the two atoms in Rydberg states respectively located at $\bf{r}$ and $\bf{r}'$ shifts the Rydberg state energy. (b)~Geometry of the system. The probe and control laser fields counter-propagate in the Rydberg gas. The involution of the weak probe pulse in the Rydberg gas is considered, while depletion of the strong control field is neglected.
(c)~Storage and retrieval of a (3+1)D light bullet, illustrated by an isosurface plot of the light intensity of the light bullet before storage ($z=0$), at the beginning of the storage ($z=5.4$\,mm) and after the storage ($z=10.8$\,mm); see text for details.}\label{fig1}
\end{figure}

We consider a three-level atomic system with a ladder-type level configuration, shown in Fig.~\ref{fig1}(a).
%
%
A weak, pulsed probe field of angular frequency $\omega_p$ (half Rabi frequency $\Omega_p$) couples the ground state $|1\rangle$ and intermediate state $|2\rangle$, and a strong, continuous-wave  control field of angular frequency $\omega_c$ (half Rabi frequency $\Omega_c$) couples state $|2\rangle$ and a Rydberg state $|3\rangle$. The probe field has a pulse length $\tau_0$ at the entrance of the medium. The electric-field vector
of the system is ${\bf E}({\bf r}, t)=\sum_{l=p,c}{\bf e}_l\, \mathcal{E}_l \,\exp [i({\bf k}_l
\cdot {\bf r}-\omega_l t)]+{\rm c.c.}$, where ${\bf e}_l$ $({\bf k}_l)$ is the unit polarization vector (wavevector) of the
electric-field component with envelope $\mathcal{E}_l\,\,(l=p,c)$. A possible experimental arrangement of beam geometry is shown in Fig.~\ref{fig1}(b).

The system works at a ultracold temperature and  the probe and control fields counter-propagate [i.e., ${\bf k}_p=(0,0,k_p)$, ${\bf k}_c=(0,0,-k_c)$], so that the center-of-mass motion of the atoms and dephasing due to the atomic collisions are negligible. Under the electric-dipole approximation, the system Hamiltonian is $\hat{H}_H={\cal N}_{a}\int {\rm d}^3 \mathbf{r} \,\hat{{\cal H}}_H(\mathbf{r},t)$ with
\begin{eqnarray} \label{eq1}
\hat{{\cal H}}_H(\mathbf{r},t)&  &=
\sum_{\alpha=1}^{3}\hbar\omega_{\alpha}\hat{S}_{\alpha\alpha}(\mathbf{r},t)-\hbar\left[\Omega_p\hat{S}_{12}(\mathbf{r},t)
+\Omega_c\hat{S}_{23}(\mathbf{r},t)\right. \nonumber\\
& & \left.+{\rm H.c.}\right] +{\cal N}_{a}\int{\rm d}^3{\mathbf{r}^\prime}\hat{S}_{33}(\mathbf{r}^\prime,t)\hbar V(\mathbf{r}^\prime-\mathbf{r})\hat{S}_{33}
(\mathbf{r},t),\nonumber
\end{eqnarray}
where $\mathcal{N}_a$ is atomic density, $\hbar \omega_{\alpha}$ is the eignenergy of atomic state $|\alpha\rangle$, $\Omega_p=(\mathbf{e}_p\cdot \mathbf{p}_{21})\mathcal{E}_p/\hbar$ and $\Omega_c=(\mathbf{e}_c\cdot \mathbf{p}_{32})\mathcal{E}_c/\hbar$ are respectively the half Rabi frequencies of the probe and control fields (with $\mathbf{p}_{\alpha\beta}$ the electric dipole matrix element associated with the transition from $|\beta\rangle$ to $|\alpha\rangle$),
$\hat{S}_{\alpha\beta}$ are transition operators $(\alpha,\beta = 1,2,3)$  satisfying the commutation relation
$\left[\hat{S}_{\alpha\beta}(\mathbf{r},t),\hat{S}_{\mu\nu}(\mathbf{r}^{\prime},t)\right]
=\left(\delta_{\alpha\nu}\hat{S}_{\mu \beta}(\mathbf{r},t)-\delta_{\mu \beta}\hat{S}_{\alpha\nu}(\mathbf{r}^{\prime},t)\right)\delta_{\mathbf{r} \mathbf{r}^{\prime}}$.
In the last term on the right hand side (RHS), $\hbar V({\bf r}-{\bf r}')=-\hbar C_6/|{\bf r}-{\bf r}'|^6$ is the van der Waals (vdW) interaction potential with  dispersion coefficient $C_6$ between the Rydberg atoms located at the positions ${\bf r}$ and ${\bf r}'$, respectively.

To be concrete, we will consider strontium atoms ($^{88}$Sr) in this work, although our approach is valid for other Rydberg atomic gases. The energy-levels shown in Fig.~1(a) are selected as $|1\rangle=|5s^2\,^1S_0\rangle$, $|2\rangle=|5s5p ^1P_1\rangle$,
$|3\rangle=|5sns^1S_0\rangle$, with $n$ the principle quantum number~\cite{Mauger}. The spontaneous emission rates of the atoms are given by
$\Gamma_{12}=2\pi\times16\,\,{\rm MHz},$ $\Gamma_{23}=2\pi\times16.7\,{\rm kHz}$, so one has $\gamma_{21}=\Gamma_{12}/2$, $\gamma_{31}=\Gamma_{23}/2$,
$\gamma_{32}=(\Gamma_{12}+\Gamma_{23})/2$. For this choice, the vdW interaction in $^1 S_0$ states is isotropically attractive ($C_6>0$), which is important to realize a self-focusing nonlinearity.

To study propagation of the probe field, we assume  the size of the atomic gas is much larger than the Rydberg blockade radius (given later). Depletion of the control field  will be neglected (except in the discussion of LB memories in Sec.~\ref{sec5}). Additionally assuming that the weak probe pulse is still a classical field (consisting of many photons),  so that a semi-classical approach can be used. With the slowly varying envelope and rotating wave approximations,
 propagation of the probe field is governed by the coupled Maxwell-Bloch equations~\cite{FIM},
\begin{equation}\label{Max}
i\left(\frac{\partial}{\partial z}+\frac{1}{c}\frac{\partial}{\partial t} \right) \Omega_{p}+\frac{c}{2\omega_p}\nabla_{\bot}^2 \Omega_p+\kappa_{12}\rho_{21}(\mathbf{r},t)=0,
\end{equation}
where $\nabla_{\bot}^2=\partial^2/\partial x^2+\partial^2/\partial y^2$,
$\kappa_{12}={\cal N}_{a}\omega_{p}|\mathbf{p}_{12}|^2/(2\epsilon_0c\hbar)$ with $\epsilon_0$ the vacuum dielectric constant. $\rho_{21}({\bf r},t)\equiv\langle {\hat S}_{21}({\bf r},t)\rangle$ is the coherence between level $|1\rangle$ and $|2\rangle$. Its dynamics is determined by the Bloch equations,
\begin{subequations} \label{eq2}
\begin{eqnarray}
&&i\frac{\partial }{\partial t}\rho_{11}-i\Gamma_{12}\rho
_{22}-\Omega
_{p}\rho_{12}+\Omega _{p}^{\ast}\rho_{21}=0,\label{eq21}\\
&&i\frac{\partial }{\partial t}\rho_{22}-i\Gamma_{23}\rho
_{33}+i\Gamma_{12}\rho _{22}+\Omega _{p}\rho_{12}-\Omega
_{p}^{\ast}\rho_{21}\nonumber\\
&&-\Omega_{c}\rho_{23}+\Omega_{c}^{\ast
}\rho_{32}=0,\label{eq22}\\
&&i\frac{\partial }{\partial t}\rho_{33}+i\Gamma_{23}\rho
_{33}+\Omega_{c}\rho_{23}-\Omega_{c}^{\ast }\rho_{32}=0,\label{eq23}\\
&&\left(i\frac{\partial }{\partial t}+d_{21}\right)
\rho_{21}-\Omega_{p}(\rho_{22}-\rho_{11})+\Omega_{c}^{\ast
}\rho_{31}=0,\label{eq24}\\
&&\left(i\frac{\partial }{\partial t}+d_{31}\right) \rho_{31}-\Omega
_{p}\rho_{32}+\Omega_{c}\rho_{21}\nonumber\\
&&-{\cal
N}_a\int{d^3\mathbf{r}^\prime V(\mathbf{r}^\prime-\mathbf{r})\rho_{33,31}(\mathbf{r}^\prime,\mathbf{r},t)}=0,\label{eq25}\\
&&\left(i\frac{\partial }{\partial t}+d_{32}\right) \rho_{32}-\Omega_{p}^{\ast}\rho_{31}-\Omega_{c}(\rho_{33}-\rho_{22})
\nonumber\\
&&-{\cal
N}_a\int{d^3\mathbf{r}^\prime V(\mathbf{r}^\prime-\mathbf{r})\rho_{33,32}(\mathbf{r}^\prime,\mathbf{r},t)}=0,\label{eq26}
\end{eqnarray}
\end{subequations}
where  $\rho_{\alpha\beta}({\bf r},t)\equiv\langle {\hat S}_{\alpha\beta}({\bf r},t)\rangle$ are the one-body correlators (one-body  density matrix elements) ~\cite{note00}, $d_{\alpha\beta}=\Delta_{\alpha}-\Delta_{\beta}+i\gamma_{\alpha\beta}$ ($\Delta_1=0$;
$\alpha, \beta= 1, 2, 3;\alpha\neq \beta)$,
$\Delta_2=\omega_p-(\omega_2-\omega_1)$ and $\Delta_3=\omega_p+\omega_c-(\omega_3-\omega_1)$ are respectively the one-photon and two-photon detunings,
$\gamma_{\alpha\beta}=(\Gamma_\alpha+\Gamma_\beta)/2+\gamma_{\alpha\beta}^{\rm col}$  with $\Gamma_\beta=\sum_{\alpha<\beta} \Gamma_{\alpha\beta}$. Here $\Gamma_{\alpha\beta}$ denotes respectively the spontaneous emission decay rate from the state $|\beta\rangle$ to the state $|\alpha\rangle$,
and $\gamma_{\alpha\beta}^{\rm col}$  represents the dephasing rate reflecting the loss of phase coherence between $|\alpha\rangle$ and $|\beta\rangle$ due to, e.g., atomic motion and the interaction between the atoms in the ground and Rydberg states.

The last terms on the left hand side of Eq.~\ref{eq2}(e) and Eq.~\ref{eq2}(f), i.e. the two-body correlators $\rho_{33,3\alpha}(\mathbf{r}^\prime,\mathbf{r},t)\equiv\langle \hat{S}_{33}
(\mathbf{r}^\prime,t)\hat{S}_{3\alpha}(\mathbf{r},t)\rangle$ ($\alpha=1,2$), result from the interaction between Rydberg atoms, where attractive (repulsive) atomic interaction lead to nonlocal self-focusing (self-defocuing) nonlinearities. Besides, there is another type of (local) nonlinearity when two-photon detuning is non-zero (i.e. $\Delta_3\neq 0$)~\cite{Bai}, contributed solely from the resonant coupling between the probe field and atoms.

From Eq.~(\ref{eq2}), we see that to solve the one-body  correlators $\rho_{\alpha\beta}({\bf r},t)$, we need to know values of the two-body correlators $\rho_{33,3\alpha}({\bf r}',{\bf r},t)\equiv \langle{\hat S}_{33}({\bf r}',t){\hat S}_{3\alpha}({\bf r},t)\rangle$ ($\alpha=1,2$), for which three-body correlators, i.e., $\rho_{\alpha\beta,\mu\nu,
\zeta\eta}({\bf r}^{\prime\prime},{\bf r}^{\prime},t)\equiv \langle{\hat S}_{\alpha\beta}({\bf r}^{\prime\prime},t){\hat S}_{\mu\nu}({\bf r}',t){\hat S}_{\zeta\eta}({\bf r},t)\rangle$, will be involved, etc.  As a result, we obtain an infinite hierarchy of equations of motion for the correlators of one-body, two-body, three-body, and so on. Therefore approximations have to be used in order to truncated the hierarchy.

\section{The (3+1)D nonlinear envelope equation}\label{sec3}

Due to the nonlinear coupling originated by the Rydberg-Rydberg interaction, it is difficult to solve the hierarchy of the equations for many-body correlators by employing conventional techniques. Fortunately, since in our consideration the probe field is relatively weak, we can solve the hierarchy of equations by using the method of multiple-scales~\cite{huang,Chen,Jeff} widely applied for solving nonlinear wave problems~\cite{Jeff}. Because our calculation is exact to third order (i.e. up to $\Omega_p^3$), higher-order $n$-body correlators ($n\ge 3$) involved in the hierarchy play no significant role and hence can be neglected.

We assume that the atoms are initially populated in state $|1\rangle$ and make the asymptotic expansion~\cite{huang,Chen}: $\rho_{\alpha1}=\sum_{l=0}\epsilon^{2l+1}\rho_{\alpha1}^{(2l+1)}$, $\rho_{32}=\sum_{l=1}\epsilon^{2l}\rho_{32}^{(2l)}$, $\rho_{\beta\beta}=\sum_{l=0}\epsilon^{2l}\rho_{\beta\beta}^{(2l)}$ with $\rho_{\beta\beta}^{(0)} =\delta_{\beta1}\delta_{\beta1}$($\alpha=2,3; \beta=1,2,3$), $\Omega_{p}=\epsilon\Omega_p^{(1)}+\epsilon^2\Omega_p^{(2)}+\epsilon^3\Omega_p^{(3)}+\cdots$, where $\epsilon$ a parameter characterizing the order of $\Omega_p$. To obtain a divergence-free expansion, all quantities on the right hand side of the expansion are considered as functions of the multiscale variables $z_l=\epsilon^{l} z$  $(l=0,1,2)$, $(x_1, y_1)=\epsilon\,(x, y)$, and $t_l=\epsilon^{l} t$ $(l=0,1)$. Substituting the above expansion into the equations of the one-body and two-body correlators (omitted here for saving space) and the Maxwell Eq.~(\ref{Max}), and comparing the coefficients of $\epsilon^l$ $(l=1,2,3...)$, we obtain a set of linear but inhomogeneous equations which can be solved order by order.

At the first order, we obtain the solution
$\Omega_p^{(1)}=F\,{\rm exp}[i(Kz_0-\omega t_0)]$, $\rho_{21}^{(1)}=(\omega+d_{31})\Omega_p^{(1)}/D(\omega)$, and $\rho_{31}^{(1)}=-\Omega_c\Omega_p^{(1)}/D(\omega)$,
with $D(\omega)=|\Omega_c|^2-(\omega+d_{21})(\omega+d_{31})$ and other $\rho_{\alpha\beta}^{(1)}$=0~\cite{note0}. In above expressions, $F=F(x_1, y_1, z_1,t_1,z_2)$  is an envelope function to be determined yet, and
$K(\omega)$ is the linear dispersion relation, given by
$K(\omega)=\omega/c+\kappa_{12}(\omega+d_{31})/D(\omega)$.
At the second order, a solvability (i.e. divergence-free) condition requires
$\partial F/\partial z_1+(1/V_g)\partial F/\partial t_1=0$,
with $V_g=(\partial K/\partial \omega)^{-1}$ the group velocity of the envelope $F$.
The explicit expression of the second-order approximate solution for the one-body  correlators, i.e. $\rho_{\alpha\beta}^{(2)}$, is presented in Appendix~\ref{ap1}.
Expansion equations for the two-body  correlators are given  in Appendix~\ref{ap2}.

With the first-order and the second-order solutions given above, we can go to the third order approximation.  A solvability condition at this order yields the (3+1)D equation for the envelope function $F$:
\begin{eqnarray} \label{NLS0}
&& i\left(\frac{\partial }{\partial
z}+\alpha_0\right)U-\frac{1}{2}K_2\frac{\partial^2U}{\partial
\tau^2}+\frac{c}{2\omega_p}\nabla_{\bot}^2 U+W_1|U|^2U  \nonumber \\
&& +\int{d^2{\bf r_\bot^\prime} G_2({\bf r_\bot^\prime}-{\bf r_\bot})|U({\bf r_\bot^\prime},z,\tau)|^2}U({\bf r}_{\bot},z,\tau)=0,
\end{eqnarray}
after returning to original variables, where ${\bf r}_{\bot}=(x,y,0)$, $U=\epsilon F$exp$(-\alpha_0 z)$ (with $\alpha_0={\rm Im}(K)$ a decay constant) $\tau=t-z/V_g$ is travelling coordinate, $K_2=\partial^{2} K/\partial \omega^{2}$ describes group velocity dispersion, $W_1=\kappa_{12}[\Omega_c^{\ast}a_{32}^{(2)}+(\omega+d_{31})(2a_{11}^{(2)}+a_{33}^{(2)})]/D(\omega)$ is the Kerr coefficient describing the {\it local} nonlinear optical response contributed from the photon-atom interaction (which vanishes if the two-photon detuning $\Delta_3$ is taken to be zero). Explicit expressions of
$a_{11}^{(2)}$, $a_{32}^{(2)}$, and $a_{33}^{(2)}$ are presented in Appendix~\ref{ap1}. $G_2({\bf r_\bot^\prime}-{\bf r_\bot})=(\kappa_{12}N_{\alpha}\Omega_c^\ast)/D(\omega)\int_{-\infty}^{+\infty}
a_{33,31}^{(3)}(\mathbf{r}^\prime-\mathbf{r})V(\mathbf{r}^\prime-\mathbf{r})dz^\prime$  (the kernel in the integral) is a reduced effective interaction potential describing the {\it nonlocal} nonlinear optical response contributed from the Rydberg-Rydberg interaction.
The expression of $a_{33,31}^{(3)}(\mathbf{r}^\prime-\mathbf{r})$ is given by
\begin{eqnarray}
& & a_{33,31}^{(3)}(\mathbf{r}^\prime-\mathbf{r}) \nonumber\\
& & \simeq\frac{-2|\Omega_c|^2\Omega_c(2\omega+d_{21}
+d_{31})/|D(\omega)|^2}{2(\omega+d_{21})|\Omega_c|^2
-D_2(\omega)[2\omega+2d_{31}-V(\mathbf{r}^\prime-\mathbf{r}) ]},
\end{eqnarray}
where $D_2(\omega)=(\omega+d_{21})(2\omega+d_{21}+d_{31})-|\Omega_c|^2$. The calculation detail for obtaining  $a_{33,31}^{(3)}(\mathbf{r}^\prime-\mathbf{r})$ is presented in Appendix~\ref{ap2}. For simplicity,
we have assumed that the spatial length of the probe pulse in the propagation (i.e., $z$) direction is much larger than the range of Rydberg-Rydberg interaction, so that a local approximation along the $z$ direction can be made~\cite{Sev}.

In Eq.~(\ref{NLS0}) the second and third terms represent respectively the dispersion and diffraction, the fourth and fifth terms are the local and nonlocal nonlinearities of the system. Because $W_1\approx 0$ when $\Delta_3=0$, it is necessary to take a non-zero $\Delta_3$ for obtaining a finite local Kerr nonlinearity. It is just the synthetic Kerr nonlinearities (i.e. the local and nonlocal ones combined together) that make the system support stable LBs, as shown below.

\section{Ultraslow weak light bullets and vortices}\label{sec4}
The form of the solution of the (3+1)D envelope Eq.~(\ref{NLS0}) depends heavily on the property of the nonlocal nonlinearity, which is characterized by the nonlocality degree of the system. The nonlocality degree can be described by using the parameter $R_b/R_0$. Here $R_b$ is the  spatial width of the effective interaction potential  $G_2({\bf r_\bot}-{\bf r}_{\bot}^{\prime})$, i.e. the Rydberg blockade radius, defined by $R_b=[|C_6 d_{21}|/(2|\Omega_c|^2)]^{1/6}$~\cite{Sev}; $R_0$ is the transverse spatial width of the probe-pulse envelope
$U$, which can be measured by the transverse beam radius of the incident probe pulse. For example, a Gaussian-type incident pulse $U=U_0\,{\rm exp}[-(r_{\bot}/R_0)^2]$, with $r_{\bot}=[x^2+y^2]^{1/2}$. Hence, the degree of nonlocality can be easily tuned by adjusting $R_0$ or $R_b$.

In the following, we take the system parameters in the dispersive regime (i.e., $|\Delta_2|\gg \Gamma_{12}$). Exact values of parameters will be given later. This allows us to divide the degree of nonlocality of the system into three typical regions~\cite{Sev,Mur}: (i)~$R_b/R_0\ll 1$ (local response region); (ii)~$R_b/R_0\sim 1$ (nonlocal response region); (iii)~$R_b/R_0\gg 1$  (strong nonlocal response region), for which different types of LB solutions can be obtained.

\subsection{Local response region}\label{sec4A}

\begin{figure}
\centering
\includegraphics[width=0.5\textwidth]{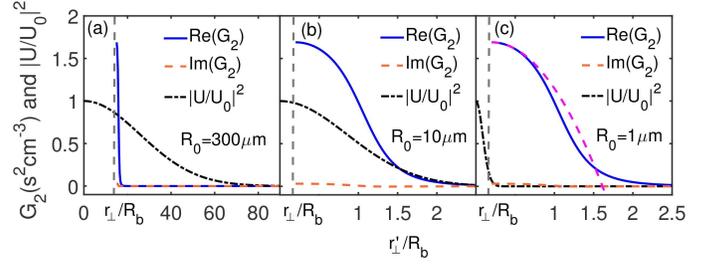}\\
\caption{\footnotesize(Color online) Effective atomic interaction potential $G_2$ as functions of $r_{\bot}^{\prime}/R_b$.  We show the local regime (a) with $R_{0}=300\,\mu{\rm m}$,
	nonlocal regime (b) with $R_{0}=10\,\mu{\rm m}$,  and strong nonlocal regime (c) with $R_{0}=1\,\mu{\rm m}$. In all situations, real parts (blue solid line) dominates over the imaginary part (orange dashed line). For a better visualization, $G_2$ has been amplified $10^8$ times.  We also show the intensity profile of the probe field $|U/U_0|^2$ (black dot-dashed line). The purple dashed line in (c) is for the function $G_2(0)+[\partial ^2G_2(0)/\partial r_\bot^2]r_\bot^2/2$. Parameters are $\Delta_2=-2\pi\times240\,{\rm MHz}$ and $R_b = [|C_6 \Delta_2| /(2|\Omega_c|^2)]^{1/6} =5.8\,\mu{\rm m}$. These parameters guarantee that  the system is in the dispersive nonlinearity regime (i.e. $|\Delta_2|\gg \Gamma_{12}$). }\label{potential}
\end{figure}

We first consider the case when the range of the Rydberg-Rydberg interaction (equivalently the spatial width of the effective atomic interaction potential $G_2$) is much smaller than the that of the beam radius of the probe pulse), i.e. $R_b/R_0\ll 1$. Fig.~\ref{potential}(a)
%
shows Re[$G_2({\bf r_\bot^\prime}-{\bf r_\bot})$] (real part of $G_2$), Im[$G_2({\bf r_\bot^\prime}-{\bf r_\bot})$] (imaginary part of $G_2$), and $|U/U_0|^2$ (the intensity profile of the probe field) as functions of $r^\prime_{\bot}/R_b$ by taking $R_{0}=300{\rm \mu m}$ (and hence $R_b/R_{0}=0.019$). From the figure we see that $|U|^2$ varies very slowly compared with $G_2$, and hence
the integral $\int{d^2\mathbf{r_\bot^\prime} G_2({\bf r}_{\bot}^{\prime}-{\bf r}_{\bot})|U(\mathbf{r_\bot^\prime},z,\tau)|^2}U(\mathbf{r},\tau)$ can be approximated by
$[\int d^2 {\bf r}_\bot^{\prime} G_2({\bf r}_{\bot}^{\prime}-{\bf r}_{\bot})]\,|U({\bf r},\tau)|^2 U(\mathbf{r},\tau)=W_2|U(\mathbf{r},\tau)|^2U(\mathbf{r},\tau)$. Here
$W_2=\int{d^2\mathbf{r_\bot^\prime} G_2(\bf{r_\bot^\prime})}=P/Q$, with
$P\simeq-i4\pi^2\kappa_{12}{\cal N}_a\sqrt{C_6}|\Omega_c|^4(2\omega+d_{21}+d_{31})$, $Q\simeq 3D(\omega)|D(\omega)|^2\sqrt{D_3(\omega)}$, and
$D_3(\omega)=D_2[2(\omega+d_{21}|\Omega_c|^2)
-D_2(2\omega+d_{31})]$. That is to say, for large probe-field beam width the nonlocal Kerr effect contributed by the Rydberg-Rydberg interaction reduces into a local Kerr nonlinearity. As a result, Eq.~(\ref{NLS0}) is simplified into
a local nonlinear Schr\"{o}dinger (NLS) equation with the Kerr coefficient given by $W_1+W_2$.

Under the EIT condition ($|\Omega_c|^2>\gamma_{21}\gamma_{31}$) and a large one-photon detuning $\Delta_2$,  imaginary parts of the coefficients in Eq.~(\ref{NLS0}) are very small.
Taking these small imaginary parts as perturbations, Eq.~(\ref{NLS0}) can be written in the dimensionless form
\begin{equation}\label{RNLSE}
i\frac{\partial u}{\partial s}-s_d\frac{\partial^2 u}{\partial \sigma^2} +2 u|u|^2
=-g_{\rm diff}\left(\frac{\partial^2}{\partial\xi^2}+\frac{\partial^2}{\partial\eta^2}\right)u+id_0 u,
\end{equation}
with $s=z/(2L_D)$, $\sigma=\tau/\tau_0$, $(\xi,\eta)=(x,y)/R_0$, $u=U/U_0$, $g_{\rm diff}=L_{\rm D}/L_{\rm diff}$, $d_0=-2L_D/L_A$, and $s_d={\rm sgn}(K_2)=\pm1$. Here $L_{\rm diff}\equiv(\omega_p R_0^2)/c$, $L_D\equiv {\tau_0}^2/|\tilde K_2|$ and $L_A\equiv 1/\alpha_0$  the typical diffraction length, dispersion length, and absorption length, respectively. Note that we have taken $L_D=L_{\rm NL}$ [with $L_{\rm NL}\equiv 1/(U_0^2\tilde{W_1}+U_0^2\tilde{W_2})$ being a typical nonlinear length], i.e., a balance of dispersion and nonlinearity is assumed to favor the formation
of solitons, thus the typical Rabi frequency of the probe field is given by $U_0\equiv (1/\tau_0)[|\tilde K_2/(\tilde W_1+\tilde W_2)|]^{1/2}$. The tilde above corresponding quantities means taking their real parts. Due to the EIT effect and large $\Delta_2$, we have $d_0\ll 1$, hence the dissipation in the system plays a negligible role; furthermore, in the present local response region the system has a large diffraction length $L_{\rm diff}$ ($\gg L_{\rm D}$) due to large $R_0$, thus we have $g_{\rm diff}\ll 1$, which means that the diffraction in the system can be neglected. Ignoring the terms on RHS of Eq.~(\ref{RNLSE})
and converting to the original variables, we obtain the bright soliton solution
\begin{equation}\label{Sol}
\Omega_p=\frac{1}{\tau_0}\sqrt{\frac{\tilde{|K_2|}}{\tilde{W}}}{\rm
sech}\left[\frac{1}{\tau_0}\left(t-\frac{z}{\tilde{V_g}}\right)
\right]e^{i\tilde{K}_0z-iz/2L_{\rm D}},
\end{equation}
if $K_2<0$ (i.e. $s_d=-1$), with $\tilde{K}_0=\tilde{K}|_{\omega=0}$.

We now give a practical example for the formation of the optical soliton given above. Choosing the parameters the same as those used in Fig.~\ref{potential}(a), we obtain the numerical values of the coefficients in Eq.~(\ref{RNLSE}), given by $K_2=(-1.03+0.06i)\times10^{-13}{\rm cm}^{-1}{\rm s}^2$, $W_1=(6.86+0.062i)\times10^{-18}{\rm cm}^{-1}{\rm s}^2$, $W_2=(1.79+0.0076i)\times10^{-14}{\rm cm}^{-1}{\rm s}^2$. We see that, as expected, the imaginary parts of these coefficients are indeed much smaller than their real parts. By taking $\tau_0=1.2\times10^{-7}$\,s, we obtain $U_0= 2\times10^7$\,s$^{-1}$, $L_D=L_{\rm NL}=1.4$\,mm, $L_A=560$\,mm, and $d_0=-0.005$, which means that the dissipation effect of the system is indeed small. In addition, because the beam radius of the probe pulse has taken to be large  ($R_0=300{\rm \mu m}$), one has $L_{\rm diff}=1226$\,mm and hence $g_{\rm diff}=0.001$, so the diffraction effect [i.e. the first term on the RHS of Eq.~(\ref{RNLSE})] in the system can indeed be neglected.

Now let's discuss properties of the soliton. With the system parameters it is easy to obtain
$\tilde{V_g}=9.8\times10^{-5}c$, i.e.
the soliton obtained has an ultraslow propagation velocity compared with $c$, the light speed in vacuum. Furthermore, the third-order nonlinear optical susceptibility, given by $\chi_{p}^{(3)}=2c|\mathbf{p}_{12}|^2(W_1+W_2)/(\hbar^2\omega_p)$, is estimated to have the value $(3.03+0.0129i)\times 10^{-8}\,{\rm m}^2\,{\rm V}^{-2}$,
which is 11 orders of magnitude higher than that obtained in conventional nonlinear optical media~\cite{Boyd}. The physical reasons for such large third-order nonlinear optical susceptibility are due to the strong Rydberg-Rydberg interaction and the EIT effect in the system. By using Poynting's vector~\cite{huang,Newell}, the maximum average power density to generate such ultraslow optical soliton is estimated to be $\bar{P}_{\rm max}=1.2\,{\rm \mu W}$. Thus, very low input power is needed for generating the optical soliton in the system.

\subsection{Nonlocal response region}\label{sec4B}
We now turn to consider nonlocal response region, corresponding to $R_b/R_0\sim 1$, which can be achieved by decreasing the probe-beam radius $R_0$ or increasing the principle quantum $n$ (thus $R_b$). One example is shown in Fig.~\ref{potential}(b). From the figure we see that $|U|^2$ varies as the same way as $G_2$. In this case, the nonlinear optical response contributed by the Rydberg-Rydberg interaction depends not only on a particular point on the light intensity, but also on a certain neighborhood of this point.

To seek possible LBs in the system, we write Eq.~(\ref{NLS0}) into the dimensionless form
\begin{eqnarray} \label{dimensionless}
&& i\frac{\partial u}{\partial s}+\left( \frac{\partial^2}{\partial \xi^2}+\frac{\partial^2}{\partial
\eta^2}\right)u+g_d\frac{\partial^2 u}{\partial \sigma^2}+g_1|u|^2u \nonumber\\
&& +\int{d\xi^\prime d\eta^\prime N(\xi-\xi^\prime,\eta-\eta^\prime)|u(\xi^\prime,\eta^\prime,s,\sigma)|^2}u(\xi,\eta,s,\sigma)
\nonumber\\
&&=id_0u,
\end{eqnarray}
through new scalings $s=z/(2L_{\rm diff})$, $(\xi,\eta)=(x,y)/R_0$, $g_d=-L_{\rm diff}\tilde{K}_2/\tau_0^2$, $d_0=-2L_{\rm diff}/L_A$, $g_1=2\tilde{W}_1|U_0|^2L_{\rm diff}$, and $N(\xi,\eta,s)=2L_{\rm diff}R_0^2|U_0|^2\tilde{G}_2(\xi,\eta)$.
Note that in Eq.~(\ref{dimensionless}) there are four main effects, i.e., the diffraction, dispersion, local nonlinearity, and the nonlocal nonlinearity (the dissipation denoted by $id_0 u$ is a small quantity). In general, for such equation
no stable LB is possible because a balance between these four effects is hard to be achieved.

However, as indicated in the first section, different from the systems considered in Refs.~\cite{Liu,Minardi,Eile,Lah,Bur,Pec1,Pec2,Gur}, stable, single LBs bounded in all spatial directions and in time can be realized in the present Rydberg-EIT system. The reasons are the following: (i)~The nonlocal optical nonlinearity coming from the Rydberg-Rydberg interaction [described by the last term on the left hand side of Eq.~(\ref{dimensionless})] has much fast response speed (with response time only in the order of $0.1\,\mu$s)~\cite{Zhang}, this is very different from the slow nonlocal optical nonlinearity in the systems studied in Refs.~\cite{Lah,Bur,Pec1,Pec2,Gur} where the response time of the nonlocal optical nonlinearity is in the range of 1\,s or even longer; (ii)~The nonlocal optical nonlinearity ($\chi_{p}^{(3)}\sim 10^{-8}{\rm m}^2{\rm V}^{-2}$)
by the Rydberg-Rydberg interaction is much stronger and possesses a faster response speed than the local optical nonlinearity ($\chi_{p}^{(3)}\sim 10^{-11}{\rm m}^2{\rm V}^{-2}$) by the photon-atom interaction [described by the fourth term on the left hand side of  Eq.~(\ref{dimensionless}), which has a response time in the order of $1\,\mu$s]. Based on these important properties (absent in the systems considered in Refs.~\cite{Liu,Minardi,Eile,Lah,Bur,Pec1,Pec2,Gur}), a single (3+1)D LB can form in the system via the following {\it two-step mechanism for self-trapping}: When a single (3+1)D probe pulse is incident into the system, it is firstly self-trapped in the two transverse (i.e. $x$, $y$) dimensions via the balance between the diffraction and the fast nonlocal optical nonlinearity; then it is further self-trapped in the longitudinal (i.e. $z$) direction by the balance between the dispersion and the slow local Kerr nonlinearity. Through such self-trapping processes, a stable single LB very different from those obtained in Refs.~\cite{Liu,Minardi,Eile,Lah,Bur,Pec1,Pec2,Gur} can be realized by the {\it synthetic} nonlocal and local optical nonlinearities in the Rydberg atomic gas. Such LB can form in a very short distance and extremely low light power due to the giant optical nonlinearities and ultraslow propagation velocity of the probe pulse resulted from the Rydberg-based EIT effect.

Before presenting the LB solution, we make an estimation on the numerical values of the coefficients in Eq.~(\ref{dimensionless}).
we choose $R_0=10$\,${\rm \mu m}$, $\tau_0=9\times10^{-7}$\,s, $U_0=3\times10^7$\,s$^{-1}$, $\Delta_2=-2\pi\times238\,{\rm MHz}$, $\Delta_3=-2\pi\times0.32\,{\rm MHz}$, ${\cal N}_a=8.2\times10^{10}$\,cm$^{-3}$ and $\Omega_c=2\pi\times16\,{\rm MHz}$. Then we obtain $R_b/R_0=0.6$, $g_d=0.134$, and $g_1=0.27$. We thus have the diffraction length $L_{\rm diff}= 1.36$\,mm and the dispersion length $L_{\rm D}(=L_{\rm LN})=10$\,mm. As expected, $L_{\rm diff}\ll L_{\rm D}$, which supports the two-step mechanism for self-trapping described above (i.e., the diffraction plays its role earlier than the dispersion).

According to the above analysis, we can assume the LB solution takes the form
%
\begin{equation} \label{ansatz}
\begin{aligned}
u=
& A(s)\,{\rm exp}\left[-\frac{\xi^2+\eta^2}{2w_s^2(s)}\right]\,{\rm sech}\left[\frac{\sigma}{w_t(s)}\right]\\
& \times {\rm exp}\left[-iC_s(s)\frac{\xi^2+\eta^2}{2w_s^2(s)}-iC_t(s)\frac{\sigma^2}{2}+i\phi(s)\right],
\end{aligned}
\end{equation}
where the parameter $w_s$ is the transverse beam width, $w_t$ is the pulse duration, $C_s$ is the wavefront curvature, and $C_t$ is the temporal chirp of the probe pulse.
All the four parameters depend on variable $s$. In the ansatz (\ref{ansatz}), the factor $\exp \left[-(\xi^2+\eta^2)/(2w_s^2(s))\right]$ is based on the balance between the diffraction and the nonlocal Kerr nonlinearity, and the factor ${\rm sech}[\sigma/w_t(s)]$ is based on the balance between the dispersion and the local Kerr nonlinearity.

\begin{figure}
\centering
\includegraphics[width=0.45\textwidth]{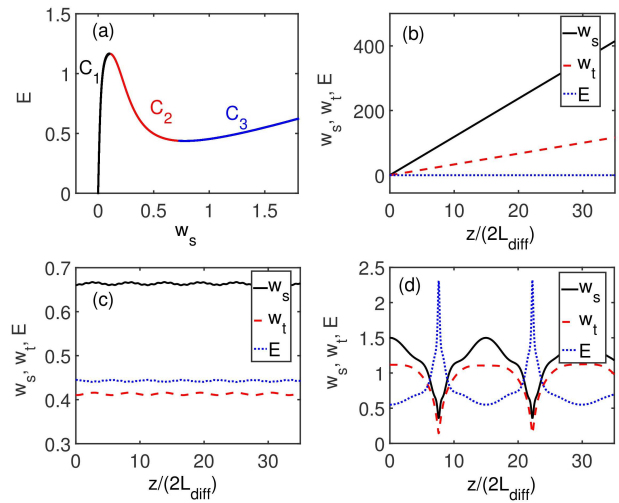}\\
\caption{(a)~LB energy $E$ as a function of the transverse beam width $w_s$. In the region where $\partial E/ \partial w_s<0$ (i.e., curve $C_2$), the LB is stable; in regions $\partial E/ \partial w_s>0$  (i.e., curves $C_1$ and $C_3$), the LB is unstable. Panels (b), (c), and (d) are numerical results of  $E$,  $w_s$ (transverse beam width), and $w_t$ (pulse duration) as a functions of $z/(2L_{\rm diff})$, obtained by choosing initial conditions from the curves $C_1$, $C_2$, and $C_3$ in the panel (a). Stability of parameter set $(w_s, w_t, E)$ with initial conditions  $(0.08, 0.06, 1.15)$ (b), $(0.66, 0.41, 0.44)$ (c) and
($(1.5, 1.1, 0.55)$ (d).}\label{stable}
\end{figure}

We employ a variational method to solve Eq.~(\ref{dimensionless}) by taking the ansatz (\ref{ansatz}) to be a LB solution~\cite{Rag}. Through a Ritz optimization procedure, the LB energy $E$, defined by $E=\int\int\int |u|^2d\xi d\eta d\sigma=2\pi A^2w_s^2w_t$, is calculated as a function of the transverse beam width $w_s$, with the result shown in Fig.~\ref{stable}(a).
We observe that there are three branches for the LB solution, i.e., curves $C_1$, $C_2$, $C_3$. In the region where $\partial E/ \partial w_s<0$ (curve $C_2$), the LB is quite stable. Yet, in regions $\partial E/ \partial w_s>0$  (curves $C_1$ and $C_3$), the LB is unstable. This conclusion is verified by linearizing variational equations around the LB solution and examining their eigenvalues and eigenfunctions; it is also checked by using a numerical simulation on Eq.~(\ref{dimensionless}) directly. Fig.~\ref{stable}(b)-Fig.~\ref{stable}(d) show results for the pulse energy $E$, the beam width $w_s$, the pulse duration $w_t$ as functions of $z/(2L_{\rm diff})$, obtained respectively by choosing initial conditions from the curve $C_1$, $C_2$, and $C_3$ in Fig.~\ref{stable}(a), i.e. $(w_s, w_t, E)=(0.08, 0.06, 1.15)$ [Fig.~\ref{stable}(b)], $(w_s, w_t, E)=(0.66, 0.41, 0.44)$ [Fig.~\ref{stable}(c)], $(w_s, w_t, E)=(1.5, 1.1, 0.55)$ [Fig.~\ref{stable}(d)]. We see that only in the case of Fig.~\ref{stable}(c), the LB's beam width $w_s$, pulse duration $w_t$, and energy $E$ keep almost unchanged, which means that the LB in the region of the curve $C_2$ is indeed stable during propagation. Note that the stable single LB solution obtained here is localized in all three spatial dimensions and also in time.

Moreover, Eq.~(\ref{dimensionless}) admits stable (3+1)D LBs carrying with orbital angular momentums (i.e., LVs). To demonstrate this, we take $u=u_{mp}(\xi,\eta,\sigma,\varphi)$ as a test solution, with
\begin{eqnarray} \label{LGmode1}
u_{mp}=
&& \frac{C_{mp}}{\sqrt{w_s}}\left[\frac{\sqrt{2}\sqrt{\xi^2+\eta^2}}{w_s}\right]^{|m|} \exp\left(-\frac{\xi^2+\eta^2}{w^2_s}\right) \nonumber \\
&& \times L_p^{|m|}\left[\frac{2(\xi^2+\eta^2)}{w^2_s}\right]{\rm sech}\left[\frac{\sigma}{w_t(s)}\right] e^{im\varphi},
\end{eqnarray}
where $L_p^{|m|}$ are the generalized Laguerre-Gauss (LG) polynomials, with $m$ and $p$ radial and azimuthal indices, respectively. The ansatz (\ref{LGmode1}) (in the absence of the factor ${\rm sech}(\sigma/w_t)$
with the normalization constant $C_{mp}=\sqrt{2^{|m|+1}p!/[\pi(p+|m|)!]}$ is called (LG)$^m_p$ mode. 
Since $\hat{L}_z\,{\rm (LG)}^m_p=m\hbar\, {\rm (LG)}^m_p$, here $\hat{L}_z=-i\hbar\partial/\partial \varphi$, (LG)$^m_p$ mode carry orbital angular momentum $m\hbar$ along the $z$ direction~\cite{Andrews}.

\begin{figure}
\centering
\includegraphics[width=0.97\columnwidth]{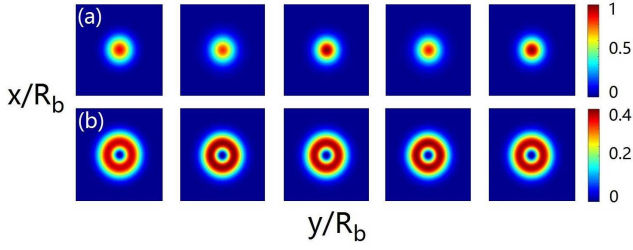}
\caption{(Color online) (a)~Evolution of $|u|^2$ with the fundamental mode (LG)$_0^0$ (i.e. light bullet) in the nonlocal response region, as a function of $x/R_b$ and $y/R_b$ when propagating to the distance respectively at $z/(2L_{\rm diff})=0$, $1$, $2$, $3$, and $4$ for atomic density ${\cal N}_a=3\times10^{10}$\,cm$^{-3}$. (b)~Evolution of $|u|^2$ with the higher order mode (LG)$_0^1$ (i.e., light vortex) for ${\cal N}_a=4.95\times10^{10}$\,cm$^{-3}$.
} \label{LGmode}
\end{figure}
%
In Fig.~\ref{LGmode} we illustrate the evolution of $|u|^2$. Panel (a) is for the LB [i.e., the fundamental mode (LG)$_0^0$] with atomic density ${\cal N}_a=3\times10^{10}$\,cm$^{-3}$; panel (b) is for the light vortex corresponding to mode (LG)$_0^1$ [${\cal N}_a=4.95\times10^{10}$\,cm$^{-3}$].
In the simulation, the beam radius $R_0$ for both the two modes is taken to be 10\,$\mu$m.
From Fig.~\ref{LGmode}(a) and Fig.~\ref{LGmode}(b), we see that for the lower-order modes (LG)$_0^0$ and (LG)$_0^1$, the pulses have nearly no deformation after propagating to $z=8L_{\rm diff}$ ($\approx 10.9$\,mm).

\begin{figure}
\centering
\includegraphics[width=0.5\textwidth]{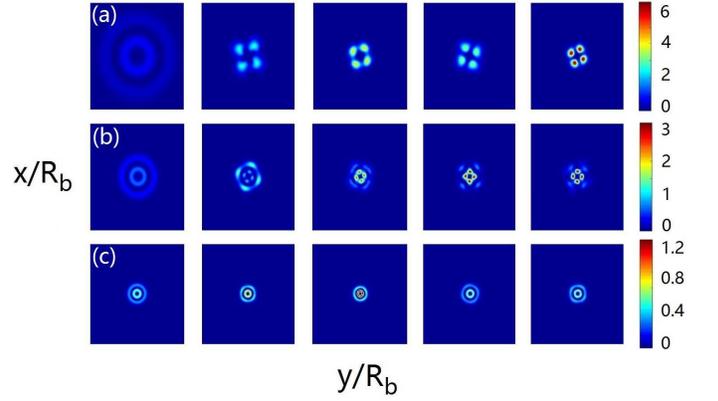}\\
\caption{(Color online) Evolution of $|u|^2$ of the light vortex corresponding to the mode (LG)$_1^2$ in the nonlocal response region, as a function of $x/R_b$ and $y/R_b$ when propagating to the distance at $z/(2L_{\rm diff})=0$, $1$, $2$, $3$, and $4$, respectively. Parameters are $R_0=10\,{\rm \mu m}$, ${\cal N}_a=9.9\times10^{10}$\,cm$^{-3}$ (a),  $R_0=5\,{\rm \mu m}$, ${\cal N}_a=1.46\times10^{11}$\,cm$^{-3}$ (b),  and $R_0=2.5\,{\rm \mu m}$, ${\cal N}_a=5.3\times10^{11}$\,cm$^{-3}$ (c). Other system parameters are the same as those used in Fig.~\ref{LGmode}.}\label{strLG}
\end{figure}

However, pulses corresponding to the higher-order LG modes may not keep its shape in the nonlocal response region. To show this, we carried out the simulation on the evolution of the LV corresponding to (LG)$_1^2$  by choosing $R_0=10\,\mu$m, $5\,\mu$m, and $2.5\,\mu$m, with the result shown in the panels (a), (b) and (c) of Fig.~\ref{strLG}, respectively.
We see that for a small beam radius [$R_0=2.5\,\mu$m; Fig.~\ref{strLG}(c)], the vortex is relatively stable compared with that having a large beam radius [$R_0=10\,\mu$m; Fig.~\ref{strLG}(a)]. This result means that if the degree of nonlocality in the system increases (i.e., $R_0$ is reduced for fixed blockade radius $R_b$), the lifetime of the vortex pulse may be increased significantly. When degree of nonlocality becomes very large, the system enters into a strongly nonlocal response region, in this case all the LVs are stable during propagation; see the discussion given below.


%

\subsection{Strongly nonlocal response region}\label{sec4C}
We lastly consider the situation when the range of the Rydberg-Rydberg interaction is much larger than that of the beam radius of the probe pulse, i.e. $R_b/R_0\gg 1$, which corresponds to the case shown in Fig.~\ref{potential}(c), where Re($G_2$), Im($G_2$), and $|U/U_0|^2$
as functions of  $r_{\bot}^{\prime}/R_b$ is shown by taking $R_{0}=1\,{\rm \mu m}$ (and hence $R_b/R_0=5.8$). We see that, compared with $|U/U_0|^2$, the response function $G_2$ is very flat near $r_{\bot}^{\prime}=r_{\bot}$, which means that $G_2({\bf r}_{\bot}^{\prime}-{\bf r}_{\bot})\approx G_2(r_\bot)\approx G_2(0)+G_2^{\prime\prime}(0)r_\bot^2/2$~\cite{snyder,Krolikowski}, plotted by the purple dashed line in Fig.~\ref{potential}(c), agreeing well with the exact one near $r_{\bot}=0$ [i.e., Re($G_2$) (blue solid line); Im($G_2$) is very small]. In this case, Eq.~(\ref{NLS0}) can be reduced into the dimensionless form
\begin{equation} \label{sd}
\begin{aligned}
&i\frac{\partial u}{\partial s}+\left(\frac{\partial^2}{\partial
\xi^2}+\frac{\partial^2}{\partial
\eta^2}\right)u+g_d\frac{\partial^2}{\partial \sigma^2}u+g_1|u|^2 u\\
&-g_4(\xi^2+\eta^2)u=id_0\,u,
\end{aligned}
\end{equation}
where $U=U_0 u\,{\rm exp}(ig_3)$, $g_{3}=2L_{\rm diff}P_0G_2(0)$, and $g_4=-L_{\rm diff}R_0^2P_0G_2^{(2)}(0)$, with $P_0=\int d^2 {\bf r}_{\bot}|U|^2$ the power of the probe pulse (approximately a constant). The definitions of $s$, $\eta$, and $\xi$ are the same as those in Sec.~\ref{sec4B}. Thus, in the strongly nonlocal response region the nonlocal Kerr nonlinearity contributed by the Rydberg-Rydberg interaction reduces into a parabolic ``external potential''. The physical reason for this reduction is that all the photons in the probe pulse experience an almost alike (effective) potential due to the very narrow probe beam radius and the very wide distribution of the potential.

Before presenting LB and LV solutions.
we first give an example of numerical values of the coefficients in Eq.~(\ref{sd}), we choose $\Delta_3=-2\pi\times48\,{\rm kHz}$, $\Delta_2=-2\pi\times160\,{\rm MHz}$, $R_0=1.86$\,${\rm \mu m}$, $\tau_0=2.6\times10^{-7}$\,s, ${\cal N}_{a}=5.8\times10^{12}$\,cm$^{-3}$, $U_0=1.5\times10^{7}$\,s$^{-1}$, and $C_6=\simeq2\pi\times167\,{\rm THz\,\mu m}^6$ (for the principle quantum number $n=120$). With these parameters, we obtain diffraction length $L_{\rm diff}=0.047$\,${\rm mm}$, and hence the blockade radius $R_b=19$\,${\rm \mu m}$\,($R_b/R_0=10.2$), $g_d=1.0-0.07i$, $g_1=2.0 - 0.014i$, $d_0=-0.03$, $g_4=2+0.1i$. Thus, the system is within the strongly nonlocal region with very small dissipation.

\begin{figure}
\centering
\includegraphics[width=0.45\textwidth]{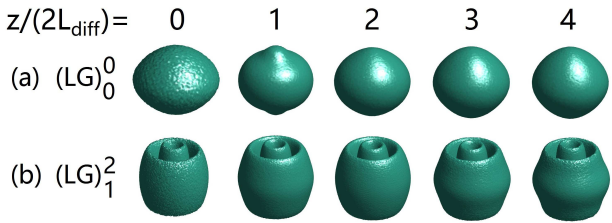}\\
\caption{(Color online) Evolution of $|u|^2$ with (LG)$_0^0$ mode [panel (a); light bullet] and (LG)$_1^2$ mode [panel (b); light vortex] in the strong nonlocal response region, obtained based on solving Eq.~(\ref{NLS0}). Isosurface plots of the pulses are shown when propagating to distances $s=z/(2L_{\rm diff}) = 0, 1, 2, 3, 4$, respectively.}\label{Iso}
\end{figure}

Similarly, a variational method is employed to solve Eq.~(\ref{sd}) by using the ansatz (\ref{LGmode1}) as test LB and LV solutions; their stability is investigated through a linear stability analysis.  Then a numerical simulation starting directly from Eq.~(\ref{dimensionless}) working in this strong nonlocal response region is carried out by taking the solutions obtained via the variational method as an initial condition, together with a random disturbance to it. Specifically, we take $U(z=0, x, y, t)=U_0\,(1+\varepsilon f_R)\,u_{mp}(z=0, x, y, t){\rm exp}(ig_3)$, with $\varepsilon$ being a typical amplitude of the perturbation, and $f_R$ being a random variable uniformly distributed in the interval [0,1]. We find that the system allows indeed stable LBs and LVs. Shown in Fig.~\ref{Iso}(a)
%
%
and Fig.~\ref{Iso}(b) are time evolution of the (3+1)D LB and LV (corresponding to the fundamental mode (LG)$_0^0$ and the higher-order mode (LG)$_1^2$) by taking $\varepsilon=0.1$, where isosurface plots are illustrated of these pulses when propagating to distances $s=z/(2L_{\rm diff}) = 0, 1, 2, 3, 4$, respectively. We observe that these (3+1)D nonlinear optical pulses relax to self-cleaned forms quite close to the unperturbed ones, and their shapes undergo no apparent change during propagation.

Based on the solution (\ref{LGmode1}) and using the parameters given above, we get the propagating velocity of the LBs and LVs
\begin{equation} \label{vg}
V_{\rm LB}\approx 1.25\times10^{-6}c,
\end{equation}
much slower than the light speed in vacuum. The maximum average power density $P_{\rm max}$ for generating such LBs and LVs can be obtained by using Poynting's vector~\cite{huang,Newell}, which is estimated to be
\begin{equation}
\bar{P}_{\rm max}\approx 0.2\, {\rm nW},
\end{equation}
which corresponds to the maximum average peak intensity $\bar{I}_{\rm max}\simeq3.6\times10^{-4}$ W cm$^{-2}$. Consequently, the (3+1)D LBs and LVs obtained in the present Rydberg-EIT system have ultraslow propagating velocity and extremely low generation power, which are very different from other schemes.

\section{Storage and retrieval of (3+1)D weak light bullets and vortices}\label{sec5}
One of main advantages of EIT is the possibility for realizing an active manipulation of optical pulses by tuning system parameters. Especially, optical pulses can be stored and retrieved through the switching off and on of the control field. In recent years, a large amount of studies on light memory by using EIT have been carried out~\cite{FIM,Chen,Novikova}, including ones performed in Rydberg atomic systems~\cite{Maxwell2013,LiL}. However, it is generally difficult to realize the memory of high-dimensional nonlinear optical pulses via conventional EIT because of the catastrophic collapse during propagation. Nevertheless, below we show that the storage and retrieval of the LBs and LVs with high efficiency and high fidelity are possible in the present Rydberg-EIT system.

To this end, we investigate the evolution of the LBs and LVs described above through solving the MB Eqs.~(\ref{eq2}) and~(\ref{Max}) by using a control field that is switched on and off adiabatically, which can be described by the switching function with the form, $\Omega_c(t)=\Omega_{c0}\{1-1/2\tanh[(t-T_{\rm off})/T_s]+1/2\tanh[t-T_{\rm on}/T_s]\}$,
where $T_{\rm{off}}$ and $T_{\rm{on}}$  are times at which the control field is of switched off and on. The duration of the switching time is $T_{s}$ and the storage time of the probe pulse is approximately given by $T_{\textrm{on}}-T_{\textrm{off}}$.

\subsection{Memory in the nonlocal response region}
We first study the storage and retrieval of the (3+1)D LB in the nonlocal response regime. The spatial waveshape of the input probe pulse is taken as a fundamental LG mode [i.e., (LG)$_0^0$ mode], and the temporal profile is assumed as a hyperbolic secant one. Fig.~\ref{fig1}(c) shows the numerical result of the probe-pulse intensity $|U/U_0|^2$ during the process of storage and retrieval. The pulse waveshapes for $z=0$ (before the storage), $z=4L_{\rm diff}$ (at the beginning of the storage), and $z=8L_{\rm diff}$ (after the storage), with $L_{\rm diff}=1.36$\,mm, are plotted. We see that when the control field $\Omega_c$ is switched on, the (3+1)D LB is created; by switching off the control field, the LB is stored in the atomic medium; then the LB is retrieved when $\Omega_c$ is switched on again. We see that the retrieved LB has nearly the same waveshape as that before the storage.

\begin{figure}
\centering
\includegraphics[width=0.38\textwidth]{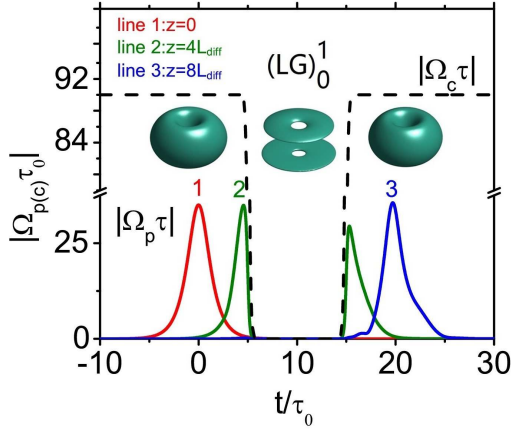}\\
\caption{(Color online) Storage and retrieval of the (3+1)D LV with the (LG)$_0^1$ mode in the nonlocal response region as a function of $t/\tau_0$. The black dashed line shows the switching-on, switching-off, and re-switching-on of the control field $|\Omega_c \tau_0|$. The curves 1, 2 and 3 are temporal profiles of the probe pulse $|\Omega_p \tau_0|$ respectively at $z= 0$ (before the storage), $z=4L_{\rm diff}$ (at the beginning of the storage), and $8L_{\rm diff}$ (after the storage), with $L_{\rm diff}=1.36\,{\rm mm}$; the corresponding isosurface plots for $|\Omega_p\tau_0|=0.1$ are also shown. In the calculation, we have set $R_0=10$\,${\rm \mu m}$, $C_6\simeq2\pi\times81.6\,{\rm GHz\,\mu m}^6$, $\Omega_p\tau_0=27$\,s$^{-1}$, $\Delta_2\tau_0=-1340$, $\Delta_3\tau_0=-1.8$, and $\Omega_{c0}\tau_0=90$ with $\tau_0=9\times10^{-7}$\,s. The control parameters are $T_{s}=0.2\tau_0$, $T_{\textrm{off}}=5\tau_0$, $T_{\textrm{on}}=15\tau_0$.
}\label{storagenonlocal}
\end{figure}

It is possible to store LVs in the nonlocal response regime in the same way.
In Fig.~\ref{storagenonlocal}
we show the storage and retrieval of a (3+1)D LV of (LG)$_0^1$ mode. The black dashed line in the figure shows the process of the switching-on, switching-off, and re-switching-on of the control field $|\Omega_c \tau_0|$. The curves 1, 2 and 3 are temporal profiles of the probe pulse $|\Omega_p \tau_0|$  when the LV propagates at $z= 0$ (before the storage), $z=4L_{\rm diff}$ (at the beginning of the storage), and $8L_{\rm diff}$ (after the storage), respectively. Corresponding isosurface plots with $|\Omega_p\tau_0|=0.1$ illustrate the storage of retrieval of the LV.

%
%

When used as optical memories, we need to know the efficiency $\eta$, which can be described by the energy ratio between the retrieved pulse and the input pulse~\cite{Novikova3}, i.e., $\eta=I_{\rm out}/I_{\rm in}$,
where $I_{\rm out}=\int_{T_{\rm on}}^{\infty}dt\iint dx dy \,|\Omega_p^{\rm out}(x,y,t)|^2$, $I_{\rm in}=\int^{T_{\rm off}}_{-\infty}dt \iint dx dy \,|\Omega_p^{\rm in}(x,y,t)|^2$, $\Omega_p^{\rm in}(x,y,t)=\Omega_p(x,y,z,t)|_{z=0}$ and $\Omega_p^{\rm out}(x,y,t)=\Omega_p(x,y,z,t)|_{z=L_z}$, with $L_z$ the medium length in the propagation direction. Based on this,
the memory efficiency of the LB (LV with (LG)$_0^1$ mode) is given by  $\eta=93.12\%$ ($\eta=92.38\%$) for $L_z=8L_{\rm diff}=10.9\,{\rm mm}$.

The quality for the preservation of pulse waveshape in the memory can be characterized by the overlap integral~\cite{Novikova3},
$J^2=I_{\rm inter}^2/(I_{\rm in}I_{\rm out})$,
where $I_{\rm inter}=|\int^{T_{\rm off}}_{-\infty}dt\iint dx dy \,\Omega_p^{\rm out}(x,y,t+\Delta T)\Omega_p^{\rm in}(x,y,t)|$, $\Delta T$ is the time interval between the peaks of the input and the retrieved probe pulses. In the case shown in Fig.~\ref{storagenonlocal}, $\Delta T=28\tau_0$, and hence we obtain  $J^2=97.26\%$ ($J^2=96.35\%$) for the LB (LV). Combined the above two parameters, one can determine the fidelity of the memory, defined by $F=\eta J^2$.
From these results, we obtain $F=90.57\%$ ($F=89.01\%$)  for the LB (LV). Thus the storage and retrieval of the LB and LV in the Rydberg atomic gas have not only high efficiency but also high fidelity.

\begin{figure}
\centering
\includegraphics[width=0.3\textwidth]{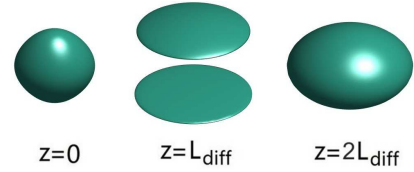}\\
\caption{(Color online) Storage and retrieval of (3+1)D LB for the case without the Rydberg-Rydberg interaction at $z= 0$ (before the storage), $z=4L_{\rm diff}$ (at the beginning of the storage), and $8L_{\rm diff}$ (after the storage), with $L_{\rm diff}=1.36\,{\rm mm}$.
}\label{withoutRyd}
\end{figure}

For comparison, Fig.~\ref{withoutRyd}
%
shows the result of the memory for the LB when the Rydberg-Rydberg interaction is absent. We see that the LB suffers a significant deformation after the storage, with the fidelity of memory only $6.3\%$ (for the memory of LVs without the Rydberg-Rydberg interaction, the fidelity is even lower), which cannot be applied for light information processing because the information is lost during the storage.

\subsection{Memory in the strong nonlocal response region}
\begin{figure}
\centering
\includegraphics[width=0.5\textwidth]{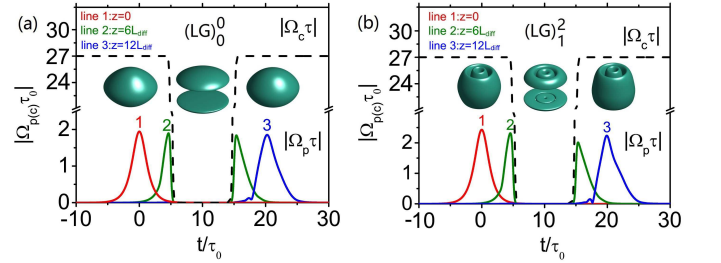}\\
\caption{(Color online) Storage and retrieval of (3+1)D LB and LV in the strong nonlocal response region as a function of $t/\tau_0$. (a)~The memory of the LB [(LG)$_0^0$) mode]. The black dashed line shows the switching on and off of the control field $|\Omega_c \tau_0|$. Curves 1, 2 and 3 are temporal profiles of the probe pulse $|\Omega_p \tau_0|$  respectively at $z= 0$ (before the storage), $z=6L_{\rm diff}$ (at the beginning of the storage), and $12L_{\rm diff}$ (after the storage), with $L_{\rm diff}=0.047\,{\rm mm}$; the corresponding isosurface plots for $|\Omega_p\tau_0|=0.1$ are also shown. (b)~The same as (a) but for the memory of the LV with the (LG)$_1^2$ mode.
}\label{storage}
\end{figure}

We now consider the storage and retrieval of the (3+1)D LBs and LVs in the strong nonlocal response region. Shown in Fig.~\ref{storage}(a)
%
[Fig.~\ref{storage}(b)]
is the numerical result of the storage and retrieval of a (3+1)D LB [LV with the (LG)$_1^2$ mode]. In the figure, the black dashed line is the time sequence of the switching-on and switching-off of the control field $|\Omega_c \tau_0|$. Curves 1, 2 and 3 are temporal profiles of the probe pulse $|\Omega_p \tau_0|$  respectively at $z= 0$ (before the storage), $z=6L_{\rm diff}$ (at the beginning of the storage), and $12L_{\rm diff}$ (after the storage), with $L_{\rm diff}=0.047\,{\rm mm}$; the corresponding isosurface plots for $|\Omega_p\tau_0|=0.1$ are also illustrated. Here system parameters used in the calculation are the same as those given in Sec.~\ref{sec4C} but with $R_0=1.86\,{\rm \mu m}$, $R_b=19\,{\rm \mu m}$, and $\tau_0=2.7\times10^{-7}$ s. From the figure we see that the LB (LV) after the storage keeps nearly the same waveshape as that before the storage.  The efficiency and fidelity of the LB memory (LV memory) in this strong nonlocal region can reach to $\eta=93.97\%$ and $F=91.86\%$  ($\eta=93.14\%$ and $F=89.11\%$). Even higher values can be reached by tuning systemic parameters (i.e., transverse beam radius $R_0$ or principle quantum $n$) to increase the degree of nonlocality.

The calculation on the storage and retrieval of other (higher-order) LVs in the strong nonlocal response region is also carried out (not illustrated here). The result shows that the memory of the higher-order LVs can also be realized and have also high efficiency and fidelity. All these results indicate that the Rydberg medium supports high-quality memory for various LG modes. Since LG modes carry definite orbital angular momentum and hence more information, the acquirement of high-quality memory by using the strong, nonlocal Rydberg-Rydberg interaction provides the possibility to realize a high-quality multi-dimensional light memory.

%


\section{Summary}\label{sec6}
In this work, we have carried out a detailed investigation on the formation, propagation, and storage of ultraslow weak-light bullets and vortices via Rydberg-EIT in a cold atomic gas. By using an approach beyond mean-field theory, we have shown that the system may acquire two types of optical nonlinearities, i.e., the giant fast-responding nonlocal optical nonlinearity (contributed by the Rydberg-Rydberg interaction) and the relatively weak, slow-responding local Kerr nonlinearity (contributed by the photon-atom interaction), both of them are enabled by the Rydberg-EIT. We have derived a (3+1)D nonlinear envelope equation governing the spatiotemporal evolution of the high-dimensional probe pulse and present various (3+1)D light bullet and vortex solutions. The light bullets and vortices obtained have very slow propagating velocity and extremely low generation power, and can be stabilized by the interplay between the synergetic local and nonlocal optical nonlinearities. These (3+1)D nonlinear optical pulses can be dynamically manipulated based on the active character of the system.  Our study expands the breadth in the study of higher dimensional, nonlocal nonlinear optics with Rydberg atomic gases. Recent experiments have reported the proof-of-principle demonstration of storing light modes with orbital angular momentum\cite{Veissier,Nicolas,Parigi,Oliveira,ding}. The highly efficient method proposed in this work has potential applications in topological quantum information processing.


\section*{Acknowledgments}

This work was supported by the NSF-China under Grants No.~11174080 and No.~11474099, by the China Postdoctoral Science Foundation under Grant No.~2017M620140, and by the Shanghai Sailing Program under Grant No.~18YF1407100. W.L. acknowledges support from the UKIERI-UGC Thematic Partnership No. IND/CONT/G/16-17/73 and EPSRC Grant No. EP/M014266/1  and EP/R04340X/1.

\appendix

\section{EXPRESSIONS OF THE SECOND-ORDER SOLUTION}\label{ap1}

Expressions of the second-order solution are given by
\begin{subequations} \label{secondorder}
\begin{eqnarray}
&& a_{21}^{(2)}=\frac{i}{\kappa_{12}}\left(\frac{1}{V_{{\rm
g}}}-\frac{1}{c}\right),\\
&& a_{31}^{(2)}=-\frac{i}{\Omega_c^{\ast}}\frac{\omega+d_{31}}{D(\omega)}
-\frac{\left(\omega+d_{21}\right)}{\Omega_c^{\ast}}a_{21}^{(2)},\\
&& a_{11}^{(2)}=\frac{[i\Gamma_{23}-2|\Omega_c|^2 M_1]M_2 -i\Gamma_{12}
|\Omega_c|^2M_3}{-\Gamma_{12}\Gamma_{23}-
i\Gamma_{12}|\Omega_c|^2 M_1},\\
&& a_{33}^{(2)}=\frac{1}{i\Gamma_{12}}\left(M_2-i\Gamma_{12}a_{11}^{(2)}
\right),\\
&&a_{32}^{(2)}=\frac{1}{d_{32}}\left(-\frac{\Omega_c}{D(\omega)}
+2\Omega_ca_{33}^{(2)}+\Omega_ca_{11}^{(2)}\right),
\end{eqnarray}
\end{subequations}
where $\rho_{\alpha1}^{(2)}=a_{\alpha1}^{(2)}\partial F/\partial t_1 {\rm exp}(i\theta)$, $\rho_{32}^{(2)}=a_{32}^{(2)}|F|^2{\rm exp}(-2\bar{\alpha}z_2)$, $\rho_{\beta\beta}^{(2)}=a_{\beta\beta}^{(2)}|F|^2{\rm exp}(-2\bar{\alpha}z_2)$($\alpha=2,3; \beta=1,2,3$) with $\theta=K(\omega)z_0-\omega t_0$,
$\bar{\alpha}=\epsilon^{-2}\alpha=\epsilon^{-2}$Im$[K(\omega)]$,
$M_1=1/d_{32}-1/d_{32}^{\ast}$, $M_2=(\omega+d_{31}^{\ast})/D(\omega)^{\ast}-(\omega+d_{31})/D(\omega)$, and $M_3=1/[D(\omega)^{\ast}d_{32}^{\ast}]-1/[D(\omega)d_{32}]$.

\section{Expansion equations of the two-body correlators}\label{ap2}
Equations of motion for the two-body correlators at second-order and third-order approximations are given in the following.

(i) {\it Second-order approximation}: we have
\begin{align}\label{twobody1}
&\begin{bmatrix}2\omega+2d_{21} & 0 & 2\Omega_c^\ast \\
    0 & 2\omega+2d_{31}-V & 2\Omega_c \\
    \Omega_c & \Omega_c^\ast & 2\omega+d_{21}+d_{31}
\end{bmatrix}
\begin{bmatrix}
\rho_{21,21}^{(2)}\\ \rho_{31,31}^{(2)}\\ \rho_{31,21}^{(2)}
\end{bmatrix}\nonumber\\
&=\begin{bmatrix}
-2\frac{\omega+d_{31}}{D(\omega)}\\0\\ \frac{\Omega_c}{D(\omega)}
\end{bmatrix}F^2e^{2i\theta},
\end{align}
\begin{align}\label{twobody2}
&\begin{bmatrix}d_{21}+d_{12} & 0 & -\Omega_c & \Omega_c^\ast\\
    -\Omega_c^\ast & \Omega_c^\ast & d_{21}+d_{13} & 0 \\
    0 & d_{31}+d_{13} & \Omega_c &-\Omega_c^\ast \\
    -\Omega_c & \Omega_c &0 & d_{21}^\ast+d_{13}^\ast
\end{bmatrix}
\begin{bmatrix}
\rho_{21,12}^{(2)}\\ \rho_{31,13}^{(2)}\\ \rho_{21,13}^{(2)} \\ \rho_{21,13}^{\ast(2)}
\end{bmatrix}\nonumber\\
&=\begin{bmatrix}
\frac{\omega+d_{31}}{D}-\frac{\omega+d_{31}^\ast}{D(\omega)^\ast}\\ \frac{\Omega_c^\ast}{D(\omega)^\ast}\\0\\ \frac{\Omega_c}{D(\omega)}
\end{bmatrix}|F|^2e^{-2\bar{\alpha} z_2}.
\end{align}
The second-order solution for two-body  correlators is given by $\rho_{\alpha1,\beta1}^{(2)}=a_{\alpha1,\beta1}^{(2)}F^2e^{2i\theta}$, $\rho_{\alpha1,1\beta}^{(2)}=a_{\alpha1,1\beta}^{(2)}|F|^2e^{-2\bar{\alpha} z_2}$ $(\alpha, \beta = 2, 3)$. Note that due to the weak probe field the vdW potential $\hbar V$ has no contribution to  $\rho_{\alpha1,1\beta}^{(2)}$, thus $\rho_{\alpha1,1\beta}^{(2)}=\rho_{\alpha1}^{(1)}\rho_{1\beta}^{(1)}$. Explicit expression of $\rho_{\alpha1,\beta1}^{(2)}$ can be directly obtained by solving Eq.~(\ref{twobody1}).

(ii) {\it Third-order approximation}. At this order, we have
\begin{widetext}
\begin{align}\label{C4}
  &\begin{bmatrix}\begin{matrix}M_{31} & \Omega_c^\ast & -i\Gamma_{23} & 0 & \Omega_c^\ast & -\Omega_c & 0 & 0 \\
    \Omega_c & M_{32} & 0 & -i\Gamma_{23} & 0 & 0 & \Omega_c^\ast & -\Omega_c \\
    0 & 0 & M_{33} & \Omega_c^\ast & -\Omega_c^\ast & \Omega_c & 0 & 0 \\
    0 & 0 & \Omega_c & M_{34} & 0 & 0 & -\Omega_c^\ast & \Omega_c \\
    \Omega_c & 0 & -\Omega_c & 0 & M_{35} & 0 & \Omega_c^\ast & 0 \\
    -\Omega_c^\ast & 0 & \Omega_c^\ast & 0 & 0 & M_{36} & 0 & \Omega_c^\ast \\
    0 & \Omega_c & 0 & -\Omega_c & \Omega_c & 0 & M_{37} & 0 \\
    0 & -\Omega_c^\ast & 0 & \Omega_c^\ast & 0 & \Omega_c & 0 & M_{38}
\end{matrix}\end{bmatrix}\begin{bmatrix}
\begin{matrix}\rho_{22,21}^{(3)}\\ \rho_{22,31}^{(3)}\\ \rho_{33,21}^{(3)}\\ \rho_{33,31}^{(3)}\\ \rho_{32,21}^{(3)}\\ \rho_{21,23}^{(3)}\\ \rho_{32,31}^{(3)}\\ \rho_{31,23}^{(3)}
\end{matrix}
\end{bmatrix}\nonumber\\
&=\begin{bmatrix}
\begin{matrix}-a_{21,12}^{(2)}+a_{21,21}^{(2)}-a_{22}^{(2)}\\
    -a_{31,12}^{(2)}+a_{21,31}^{(2)}\\-a_{33}^{(2)}\\0\\ a_{21,31}^{(2)}-a_{32}^{(2)}\\-a_{32}^{\ast(2)}-a_{21,13}^{(2)}\\ a_{31,31}^{(2)}\\
    -a_{31,13}^{(2)}
\end{matrix}
\end{bmatrix}|F({\bf r^\prime})|^2F({\bf r})e^{-2\bar{\alpha} z_2^\prime}e^{i\theta},
\end{align}
\end{widetext}
where $M_{31}=\omega+i\Gamma_{12}+d_{21}$, $M_{32}=\omega+i\Gamma_{12}+d_{31}$, $M_{33}=\omega+i\Gamma_{23}+d_{21}$, $M_{34}=\omega+d_{31}+i\Gamma_{23}-V$, $M_{35}=\omega+d_{32}+d_{21}$, $M_{36}=\omega+d_{23}+d_{21}$, $M_{37}=\omega+d_{32}+d_{31}-V$ and $M_{38}=\omega+d_{23}+d_{31}$. Form these equations
we obtain the third order solution
\begin{equation}
\rho_{33,31}^{(3)}=a_{33,31}^{(3)}|F({\bf r^\prime})|^2F({\bf r})e^{-2\bar{\alpha} z_2^\prime}e^{i\theta},
\end{equation}
with
\begin{equation}\label{order33313}
\begin{aligned}
a_{33,31}^{(3)}&=\frac{P_0+P_1V(\mathbf{r}^\prime-\mathbf{r})+P_2V(\mathbf{r}^\prime-\mathbf{r})^2}{Q_0
+Q_1V(\mathbf{r}^\prime-\mathbf{r})+Q_2V(\mathbf{r}^\prime-\mathbf{r})^2+Q_3V(\mathbf{r}^\prime-\mathbf{r})^3},\\
&\simeq\frac{-2|\Omega_c|^2\Omega_c(2\omega+d_{21}+d_{31})/|D(\omega)|^2}{2(\omega+d_{21})|\Omega_c|^2-D_2(\omega)[2\omega+2d_{31}-V(\mathbf{r}^\prime-\mathbf{r}) ]}.
\end{aligned}
\end{equation}
Here $P_n$ and $Q_n~(n=0,1,2,3)$ are functions of the spontaneous emission decay rate $\gamma_{\mu\nu}$, detunings $\Delta_\mu$, and half Rabi frequency $\Omega_c$.



\begin{references}
\bibitem{note0}The frequency and wave number of the probe field are given by
$\omega_p+\omega$ and $k_p+K(\omega),$ respectively. Thus
$\omega=0$ corresponds to the center frequency of the probe field.
\end{references}
\end{document}